\documentstyle[aps,twocolumn,epsf,rotate]{revtex}
\newcommand{\nwc}{\newcommand}
\nwc{\cl}  {\clubsuit}
\nwc{\ds}  {\displaystyle}
\nwc{\ra}  {\rightarrow}
\nwc{\ul}  {\underline}
\nwc{\nnn} {\nonumber \vspace{.2cm} \\ }
\nwc{\Dc}  {{\cal D}}
\nwc{\Th} {\Theta}
\nwc{\th} {\theta}
\nwc{\vth} {\vartheta}
\nwc{\eps}{\epsilon}
\nwc{\si} {\sigma}
\nwc{\Gm} {\Gamma}
\nwc{\gm} {\gamma}
\nwc{\bt} {\beta}
\nwc{\La} {\Lambda}
\nwc{\la} {\lambda}
\nwc{\om} {\omega}
\nwc{\Om} {\Omega}
\nwc{\dt} {\delta}
\nwc{\Si} {\Sigma}
\nwc{\Dt} {\Delta}
\nwc{\al} {\alpha}
\nwc{\vph}{\varphi}
\nwc{\zt} {\zeta}
\def\tr{\mathop{\rm tr}}
\def\Tr{\mathop{\rm Tr}}

\def\VEV#1{\left\langle #1\right\rangle}

\nwc{\Id}  {{\bf 1}}
\nwc{\sgn}  {{\rm sgn}}
\nwc{\diag} {{\rm diag}}
\def\slash#1{#1\!\!\!/\!\,\,}

\def\MeV {\,{\rm  MeV}}

\def \lta {\mathrel{\vcenter
     {\hbox{$<$}\nointerlineskip\hbox{$\sim$}}}}
 
\def\KK{{\rm I\kern -.2em  K}}
\def\NN{{\rm I\kern -.16em N}}
\def\RR{{\rm I\kern -.2em  R}}
\def\ZZ{Z \kern -.43em Z}
\def\QQ{{\rm \kern .25em
             \vrule height1.4ex depth-.12ex width.06em\kern-.31em Q}}
\def\CC{{\rm \kern .25em
             \vrule height1.4ex depth-.12ex width.06em\kern-.31em C}}
\def\ZZZ{Z\kern -0.31em Z}

\begin{document}

\title{
  \vskip-1.0cm{\baselineskip14pt}
  \centerline{\normalsize\hskip12.5cm HD--THEP 98--39}
  \vskip0.5cm
  The chiral phase transition from the exact RG}

\author{Dirk--Uwe Jungnickel\thanks{Email:
    D.Jungnickel@thphys.uni-heidelberg.de}}

\address{Institut f\"ur Theoretische Physik, Universit\"at Heidelberg,
         D--69120 Heidelberg, Germany}

\date{Talk given at the 5th International
Workshop on Thermal Field Theories\\ and Their Applications, Regensburg,
10--14 August 1998}

\maketitle

\begin{abstract}
  A brief introduction is given to the concept of the effective average
  action. Its dependence on the averaging or coarse graining scale is
  governed by an exact RG equation for which nonperturbative approximation
  schemes are described. This formalism is applied to the computation of
  the equation of state for two flavor QCD within an effective linear quark
  meson model.  Our results allow to derive the temperature and quark mass
  dependence of quantities like the chiral condensate or the pion mass. A
  precision estimate of the universal critical equation of state for the
  three--dimensional $O(4)$ Heisenberg model is given.  The exact RG
  formalism applied to the linear quark meson model is demonstrated to
  provide an explicit link between the $O(4)$ universal behavior near the
  critical temperature and zero quark mass on one hand and the physics at
  low temperatures and realistic current quark masses, i.e., the domain of
  validity of chiral perturbation theory on the other hand.
\end{abstract}

\narrowtext


\section{The chiral transition and the linear quark meson model}
\label{sec1}

The chiral phase transition is very difficult to tackle analytically. The
main obstacles which arise are twofold. QCD as the microscopic theory of
strong interactions is formulated in terms of quarks and gluons.  The IR
behavior of strong interactions is, however, dominated by collective degrees
of freedom like mesons. For instance, the relevant modes for the chirally
broken phase are Goldstone bosons which, in QCD, are quark--antiquark bound
states.  Furthermore, for the relevant length scales the running gauge
coupling $\al_s$ is large and a perturbative treatment is questionable.

A popular way to circumvent the first difficulty is the use of effective
field theories for the most important degrees of freedom. The most
prominent example is chiral perturbation theory based on the nonlinear
sigma model \cite{Wei77-1,GL82-1,GL85-1} which describes the IR behavior of
QCD in terms of the Goldstone bosons of spontaneous chiral symmetry
breaking. This yields a very successful effective formulation of strong
interactions dynamics for momentum scales up to several hundred $\MeV$ or
temperatures of several $10\MeV$.  For somewhat higher scales additional
degrees of freedom like the sigma meson or the light quark flavors will
become important and should be included explicitly.  We will therefore
rather work with an effective linear quark meson model
\cite{JW95-1,JW98-1,JW96-1,JW98-2}. Here the lightest mesonic degrees of
freedom are encoded in a complex scalar field matrix $\Phi$ which can be
thought of as a (color--neutral) quark--antiquark composite, $\Phi^{a
  b}\sim\overline{\Psi}_L^b\Psi_R^a$, where $a,b$ label the $N_f$ light
quark flavors. Spontaneous chiral symmetry breaking corresponds to a scalar
vacuum expectation value $\VEV{\Phi^{a b}}=\overline{\si}_0\dt^{a b}$ with
$\overline{\si}_0\neq0$.  For the scale--dependent effective action which
describes the interactions of the light scalar and pseudoscalar mesons with
quarks we make the Ansatz
\begin{eqnarray}
 \label{GammaEffective}
 \ds{\Gm_k[\Phi,\Phi^\dagger,\Psi,\overline{\Psi}]} &=& \ds{
   \int d^4x\Bigg\{
   Z_\Phi\tr\left[\partial_\mu\Phi^\dagger\partial^\mu\Phi\right]
   }\nnn
 && \ds{\hspace{-2cm}+
   Z_\Psi\overline{\Psi}_a i\slash{\partial}\Psi^a+
   U_k(\Phi,\Phi^\dagger)-
   \frac{1}{2}\tr\left(
     \Phi^\dagger\jmath+\jmath^\dagger\Phi\right)
   }\\[2mm]
 && \ds{\hspace{-2cm}+
   \overline{h}\,\overline{\Psi}^a\left(\frac{1+\gm_5}{2}\Phi_{ab}-
     \frac{1-\gm_5}{2}(\Phi^\dagger)_{ab}\right)\Psi^b
     \Bigg\}\nonumber
   }
\end{eqnarray}
where $k$ denotes the relevant momentum scale in a way defined more
precisely below.  The meson and quark wave function renormalizations,
$Z_\Phi$ and $Z_\Psi$, respectively, are now functions of $k$ and
$\overline{h}$ is a $k$--dependent Yukawa coupling. The scale--dependent
effective potential is denoted as $U_k$.

We will assume that (\ref{GammaEffective}) describes the most important low
energy degrees of freedom of strong interactions for scales $k$ below a
``compositeness scale'' $k_\Phi$ at which the original gauge interaction of
QCD becomes strong enough to trigger the formation of the light mesonic
bound states. The effect of explicit chiral symmetry breaking due to
non--vanishing current quark masses is represented by an external source
$\jmath$ for the scalar meson field $\Phi$. It is related to the average
current quark mass $\hat{m}=(m_u+m_d)/2$ (we neglect here isospin violating
effects) by~\cite{BJW98-1}
\begin{equation}
  \label{eq:aaa0001}
  \jmath=2\overline{m}_{k_\Phi}^2\hat{m}
\end{equation}
where $\overline{m}_{k_\Phi}$ denotes the bare scalar mass parameter
contained in $U_k$ at the compositeness scale $k_\Phi$.  The main
approximation made in (\ref{GammaEffective}) is the neglect of higher
derivative terms as well as higher dimensional interactions among quarks
and mesons.  We note that this model is a generalization of the
Nambu--Jona--Lasinio (NJL) model \cite{NJL61-1,Bij96-1} where the
four--fermion interaction has been eliminated in favor of an auxiliary
field $\Phi$. Beyond the NJL--model we allow here for a non--vanishing
kinetic term for the scalar mesons, i.e., $Z_\Phi\neq0$, as well as an
arbitrary form of the effective average potential $U_k$ for all values of
$k$.

The success of the NJL--model to describe the strong chiral dynamics at low
energies (see, e.g., \cite{Bij96-1} and references therein) serves as a
further support for the use of (\ref{GammaEffective}). Our motivation to
consider the linear quark meson model as an effective field theory for the
IR behavior of QCD is, however, not only limited to the relation
of~(\ref{GammaEffective}) to NJL--models. It can be demonstrated that,
similarly to chiral perturbation theory, also within the scalar part of the
linear quark meson model a systematic expansion of observables like meson
masses or decay constants in powers of the light current quark masses is
possible~\cite{JW98-1}. This allows for a successful ``prediction'' of the
light quark mass ratios~\cite{JW96-1} and a computation of several of the
$L_i$ coefficients of higher dimensional operators of chiral perturbation
theory~\cite{JW98-2} in good agreement with the results of other methods.

We will restrict ourselves here to the two lightest quark flavors and
neglect isospin violation. Furthermore, we decouple the $\eta^\prime$ and
the $a_0$ isotriplet.  This is justified by their relatively large masses
and can be achieved for $N_f=2$ in a chirally invariant manner.  This
leaves us with the $O(4)$--symmetric Gell--Mann--Levy model for the three
pions and the sigma meson coupled here, however, to the up and the down
quarks (instead of the nucleons). It should be noted, though, that our
justification for the decoupling of the $\eta^\prime$ meson is less clear
at finite temperature.  It has been speculated~\cite{PW84-1,Shu94-1} that
close to the chiral transition temperature an effective restoration of the
axial $U_A(1)$ symmetry might take place. In this case the $\eta^\prime$
would become degenerate with the pions even in the spontaneously broken
phase. In principle this question could be address within the linear quark
meson model by allowing for a finite explicitly $U_A(1)$ breaking scalar
interaction term $\sim\overline{\nu}(\det\Phi+\det\Phi^\dagger)$ in the
effective average potential $U_k$. This would lead to a finite
$\eta^\prime$ mass even in the chiral limit and the temperature dependence
of the coupling $\overline{\nu}(T)$ would yield the require information
about an approximate restoration of $U_A(1)$ within the linear quark meson
model. We leave this generalization of our model to future work.

We furthermore note that even within this relatively simple effective model
we have to deal with strong couplings. For instance, the renormalized
Yukawa coupling acquires a value $h\simeq6$ to reproduce a realistic
constituent quark mass of $M_q=h f_\pi/2\simeq300\MeV$. Thus a
nonperturbative method is required which we discuss next.

\section{Effective average action and exact RG} 
\label{sec2}

The concept of the effective average action $\Gm_k$ is most easily
introduced by considering an $O(N)$--symmetric scalar model with real
fields $\chi^a$ in $d$ Euclidean dimensions and classical action $S[\chi]$.
We define the scale dependent generating functional for connected Green
functions as
\begin{equation}
  \label{AAA001}
  W_k[J]=\ln\int\Dc\chi\exp\left\{-S_k[\chi]+\int d^d x
  J_a(x)\chi^a(x)\right\}
\end{equation}
with IR cutoff scale $k$. Here $S_k[\chi] = S[\chi]+\Dt S_k[\chi]$ with
$S[\chi]$ the classical action and
\begin{equation}
  \Dt S_k[\chi] = \frac{1}{2}\int\frac{d^d q}{(2\pi)^d}
  R_k(q^2)\chi_a(-q)\chi^a(q)\; .
\end{equation}
We require that the IR cutoff function $R_k(q^2)$ vanishes rapidly for
$q^2\gg k^2$ whereas for $q^2\ll k^2$ it behaves as $R_k(q^2)\simeq k^2$.
This implies that all Fourier modes $\chi^a(q)$ with momenta smaller than
$k$ acquire an effective mass $\sim k$ and decouple while the high momentum
modes of $\chi^a$ are not affected by $R_k$.  The classical fields
$\Phi^a\equiv\VEV{\chi^a}=\dt W_k[J]/\dt J_a$ now depend on $k$, and the
effective average action is defined as
\begin{equation}
  \label{AAA60}
  \Gm_k[\Phi]=-W_k[J]+\int d^dx J_a(x)\Phi^a(x)- 
  \Delta S_k[\Phi] \; .
\end{equation}
In order to define a reasonable coarse grained free energy we have subtracted
in (\ref{AAA60}) the infrared cutoff piece. This guarantees that the only
difference between $\Gamma_k$ and $\Gamma$ is the effective IR cutoff in the
fluctuations. Furthermore, it has the consequence that $\Gamma_k$ does not
need to be convex for $k>0$ whereas a pure Legendre transform is always
convex. (The coarse grained free energy becomes convex only for $k\ra0$.)
This is important for the description of phase transitions, in particular,
first order ones. We note that the effective average action is a continuum
implementation of the Wilson--Kadanoff block--spin
action~\cite{Wil71-1,Kad66-1}, however, formulated here for the generating
functional of $1PI$ Green functions.

For choices of the IR cutoff function $R_k$ like
\begin{equation}
  \label{AAA61}
  R_k(q^2)=\frac{Z_\Phi q^2}
  {e^{q^2/k^2}-1}
\end{equation}
the effective average action $\Gm_k$ interpolates between the classical
action in the ultraviolet and the full effective action in the infrared:
\begin{eqnarray}
  \label{hhh001}
  \ds{\lim_{k\ra\infty}\Gm_k[\Phi]} &=& \ds{
    S[\Phi]
    }\nnn
  \ds{\lim_{k\ra0}\Gm_k[\Phi]} &=& \ds{
    \Gm[\Phi]
    }
\end{eqnarray}
The high momentum modes are very effectively integrated out in
(\ref{AAA001}) because of the exponential decay of $R_k$ for $q^2\gg k^2$.
All symmetries of the model that are respected by the IR cutoff $\Delta_k
S$ are automatically symmetries of $\Gamma_k$. Hence there is no problem
incorporating chiral fermions since a chirally invariant cutoff can be
formulated~\cite{Wet90-1,BW93-1,JW96-1}.

The dependence of $\Gamma_k$ on the coarse graining scale $k$ is governed
by an exact renormalization group (RG) equation~\cite{Wet93-3}
\begin{equation}
  \partial_t\Gm_k[\Phi] = \frac{1}{2}\Tr\left\{
  \frac{1}
  {\Gm_k^{(2)}[\Phi]+R_k}
  \partial_t R_k\right\} \; .
  \label{ERGE}
\end{equation}
Here $t=\ln(k/\La)$ with some arbitrary momentum scale $\La$. For the
linear quark meson model we will identify $\Lambda$ for convenience with
the compositeness scale $k_\Phi$ where the mesons a thought to form within
QCD.  The trace includes a momentum integration as well as a summation over
internal indices. The second functional derivative
\begin{equation}
 \label{AAA69}
 \left[\Gm_k^{(2)}\right]_{ab}(q,q^\prime)=
 \frac{\dt^2\Gm_k}{\dt\Phi^a(-q)\dt\Phi^b(q^\prime)}
\end{equation}
denotes the {\em exact} inverse average propagator.  The exact RG
equation~(\ref{ERGE}) allows to start with a microscopic (classical or
possibly already effective) action at some UV scale $\La$ and successively
include all quantum fluctuations by lowering $k$ to zero where subsequently
all physics is extracted.

Eq.~(\ref{ERGE}) is a partial functional differential equation which, in
general, will be difficult to solve. Its usefulness depends crucially on
the existence of systematic approximation schemes.  Besides conventional
perturbative or large--$N$ expansions there are two main directions which
have been explored in the literature: an expansion of the effective
Lagrangian in powers of derivatives
\begin{eqnarray}
  \label{DerExp}
  \ds{\Gamma_k[\Phi]} &=& \ds{
    \int d^d x\Big\{
    U_k(\rho)+\frac{1}{2}Z_{k}(\Phi)
    \partial_\mu\Phi^a\partial^\mu\Phi_a
    }\nnn
  &+& \ds{
    \frac{1}{4}Y_{k}(\rho)\partial_\mu\rho
    \partial^\mu\rho+
    O(\partial^{4})\Big\}
    }
\end{eqnarray}
where $\rho\equiv\frac{1}{2}\Phi_a\Phi^a$, or one in powers of the fields
\begin{eqnarray}
  \label{FieldExp}
  \ds{\Gamma_k[\Phi]} &=& \ds{
    \sum_{n=0}^\infty\frac{1}{n!}\int
    \left(\prod_{j=0}^n d^d x_j
      \left[\Phi(x_j)-\Phi_0\right]\right)
        }\nnn
  &\times& \ds{
    \Gamma_k^{(n)}(x_1,\ldots,x_n)
    }\; .
\end{eqnarray}
(If one chooses $\Phi_0$ as the $k$--dependent VEV of $\Phi$, the series
(\ref{FieldExp}) starts effectively at $n=2$.) The flow equations for the
$1PI$ $n$--point functions $\Gamma_k^{(n)}$ are obtained by functional
differentiation of eq.~(\ref{ERGE}).  The formation of mesonic bound
states~\cite{EW94-1}, which typically appear as poles in the (Minkowskian)
four--quark Green function, is most efficiently described by an expansion
like (\ref{FieldExp}).  On the other hand, a parameterization of $\Gamma_k$
as in (\ref{DerExp}) seems particularly suited for the study of phase
transitions and will therefore be our choice here.

Our Ansatz for the effective average action of the linear quark meson model
is provided by (\ref{GammaEffective}).  Inserting it into the exact RG
equation~(\ref{ERGE}) this complicated functional differential equation is
converted into a set of coupled ordinary and partial differential equations
for the $k$--dependence of the two wave function renormalizations $Z_\Phi$,
$Z_\Psi$, the Yukawa coupling $\overline{h}$ and the effective average
potential $U_k$. For the sake of brevity we will not display here the full
set of flow equations and refer the interested reader
to~\cite{JW96-1,BJW98-1}. It is, however, illustrative to consider the flow
equation for the effective average potential as an example.  With the
abbreviations $u(t,\tilde{\rho})=k^{-d} U_k(\rho)$ where
$\tilde{\rho}=k^{2-d}Z_\Phi\rho$ it is given by
\begin{eqnarray}
  \label{eq:3034}
    \ds{\frac{\partial}{\partial t}u} &=& \ds{
      -d u+\left(d-2+\eta_\Phi\right)
      \tilde{\rho}u^\prime+
      2v_d\Big\{
      3l_0^d(u^\prime;\eta_\Phi)
      }\nnn
    && \ds{\hspace{-5mm}+
      l_0^d(u^\prime+2\tilde{\rho}u^{\prime\prime};\eta_\Phi)-
      2^{\frac{d}{2}+1}N_c
      l_0^{(F)d}(\frac{1}{2}\tilde{\rho}h^2;\eta_\Psi)
      \Big\} }\; .
\end{eqnarray}
Here $v_d^{-1}\equiv2^{d+1}\pi^{d/2}\Gm(d/2)$ and primes denote derivatives
with respect to $\tilde{\rho}$.  We will always use in the following
$N_c=3$ for the number of quark colors and $d=4$. Eq.~(\ref{eq:3034}) is a
partial differential equation which governs the flow of the effective
average potential $u(t,\tilde{\rho})$ with $k$ starting from a fixed UV
scale to $k=0$. Supplemented by similar (but ordinary) differential
equations for the flow of the remaining parameters
$\eta_\Phi\equiv-\frac{d}{d t}\ln Z_\Phi$, $\eta_\Psi\equiv-\frac{d}{d
  t}\ln Z_\Psi$ and $h^2\equiv Z_\Phi^{-1}Z_\Psi^{-2}\overline{h}$ it can
be solved numerically on a computer. 

An important nonperturbative ingredient appearing in all these flow
equations and, in particular, also in~(\ref{eq:3034}) are so called mass
threshold functions. Full definitions of all threshold functions relevant
for the truncation~(\ref{GammaEffective}) of the linear quark meson model
can be found in~\cite{JW95-1}. A typical example appearing
in~(\ref{eq:3034}) is
\begin{equation}
 \label{AAA85}
 l_n^4(\frac{M^2}{k^2};\eta_\Phi)=8n\pi^2
 k^{2n-4}\int\frac{d^4q}{(2\pi)^4}
 \frac{\partial_t(Z_\Phi^{-1}R_k(q^2))}
 {\left[ P(q^2)+M^2\right]^{n+1}}
\end{equation}
with $P(q^2)=q^2+Z_\Phi^{-1}R_k(q^2)$. These functions decrease
monotonically with $M/k$ and decay $\sim(k^2/M^2)^{n+1}$ for $k\ll M$ where
$M$ can be seen from~(\ref{eq:3034}) to be a renormalized scalar mass of
the model.  This implies that the main effect of the threshold functions is
to cut off quantum fluctuations of particles with masses $M^2\gg k^2$.
These functions therefore automatically and smoothly decouple massive modes
from the evolution of the system whenever $k$ becomes smaller than the
($k$--dependent) mass of a particle. The relevant masses typically receive
contributions from arbitrarily high powers of the coupling constants of the
theory and are therefore nonperturbative in nature. In fact, this
nonperturbative dependence of the flow equations on the coupling constants
of the model is already manifest in~(\ref{ERGE}).  The second functional
derivative $\Gm_k^{(2)}$ generically depends on all couplings present in
the effective average action which therefore appear in denominators of the
beta functions for these very same couplings.  This nonperturbative
dependence of the beta functions on the couplings is an important
difference to Polchinski's exact RG equation~\cite{Pol84-1}.  It
corresponds to a resummation of the contributions from infinitely many
Feynman diagrams already at the level of the beta functions. It is the main
reason that nonperturbative problems like the computation of critical
exponents (see below) can be carried out in a straightforward way.

\section{Initial conditions for the RG flow}
\label{sec3}

We will assume that the linear quark meson model is a reasonable effective
description of the chiral IR dynamics of QCD for scales $k$ below a
``compositeness scale'' $k_\Phi$ where the mesons are assumed to form due
to strong gluonic interactions. For a solution of our coupled set of RG
equations for $Z_\Phi$, $Z_\Psi$, $\overline{h}$ and $U_k$ we therefore
need initial conditions at the scale $\La=k_\Phi$. We motivate these
conditions by exploiting the relation of the linear quark meson model to
the NJL--model. There the scalar meson field $\Phi$ is introduced as an
auxiliary field for $k=k_\Phi$. This implies that its wave function
renormalization $Z_\Phi$ vanishes at the compositeness scale.  Furthermore,
in the NJL--limit the effective average potential $U_k$ is a purely
quadratic function of $\Phi$ with a positive mass squared coefficient
$\overline{m}_{k_\Phi}^2\sim1/G$ where $G$ is the four--fermi coupling of
the NJL--model. We will be somewhat more general here and allow for a
non--vanishing by small scalar wave function renormalization
$Z_\Phi(k_\Phi)\ll1$ at the compositeness scale. Moreover, we will assume
that the effective average potential is almost quadratic for $k=k_\Phi$.
More precisely, we take a starting function $U_{k_\Phi}(\Phi,\Phi^\dagger)$
which has its absolute minimum at the origin $\Phi=\Phi^\dagger=0$ but is
otherwise unrestricted.  The two remaining initial conditions can be freely
chosen as $Z_\Psi(k_\Phi)=\overline{h}(k_\Phi)=1$. The condition on the
effective average potential implies that the evolution of the system starts
at $k_\Phi$ in the chirally symmetric regime.  Clearly, because of quark
fluctuations with a large renormalized Yukawa coupling, the scalar mass
squared term will decrease with $k$ and finally turn negative, therefore
spontaneously breaking chiral symmetry.  This typically happens at scales
$k_{\chi SB}\simeq400\MeV$ well above $\Lambda_{\rm QCD}$. This may serve
as a qualitative justification for the neglect of confinement effects in
this as well as in the limiting NJL--model: The only degrees of freedom in
the model which experience confinement directly are the two light quark
flavors. Yet, they acquire a constituent mass of the same order as $k_{\chi
  SB}$ before $k$ gets too close to $\La_{\rm QCD}$.  They will therefore
decouple from the evolution of the mesonic system before confinement
effects start to modify the quark propagator significantly.

The smallness of $Z_\Phi(k_\Phi)$ implies that all mesons have large
renormalized masses near $k_\Phi$ and therefore decouple effectively. The
model is thus dominated by quark fluctuations for the beginning of its
$k$--evolution. As a consequence, the flow equations exhibit an approximate
partial IR fixed point not far below $k_\Phi$~\cite{JW96-1,BJW98-1}.  More
precisely, except for the meson mass term all scalar self interactions
contained in $U_k$ divided by appropriate powers of the renormalized Yukawa
coupling rapidly approach fixed points and almost all information about the
initial conditions at $k_\Phi$ is washed out. This implies an enormous
predictive power of the exact RG analysis of the linear quark meson model
which goes far beyond the chiral Ward identities and has been successfully
tested~\cite{JW96-1,BJW98-1}.  There are only three unknown parameters left
in our model which have to be fixed using phenomenological input: the
(bare) scalar mass $\overline{m}_{k_\Phi}$ at $k_\Phi$ as the only relevant
initial condition, the compositeness scale $k_\Phi$ itself and the average
light current quark mass $\hat{m}$.  Using $f_\pi=93\MeV$, $M_\pi=135\MeV$
and a constituent mass $M_q=300\MeV$ as input we find e.g.,
$k_\Phi\simeq600\MeV$.

\section{Chiral phase transition at non--vanishing temperature}
\label{sec4}

\begin{figure}
  \unitlength1.0cm
  \begin{center}
    \begin{picture}(8.5,7.0)
      \put(0.5,0.5){
        \epsfysize=7.5cm
        \epsfxsize=6.0cm
        \rotate[r]{\epsfbox{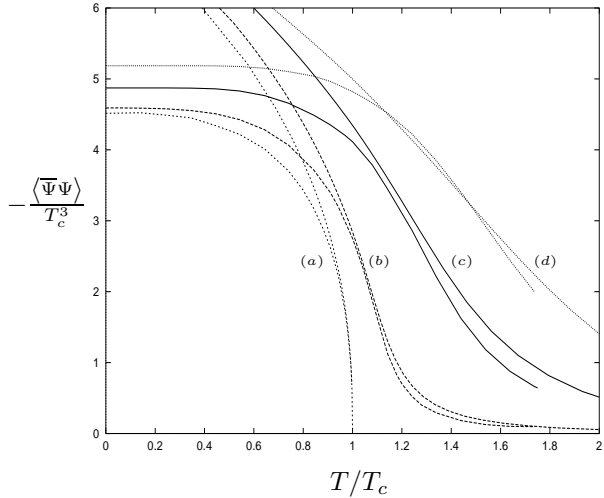}}
        }
      \put(4.3,0.0){\bf $T/T_{c}$}
      \put(0.0,3.7){\bf $-\frac{\VEV{\overline{\Psi}\Psi}}{T_{c}^3}$}
      \put(3.9,3.0){\tiny $(a)$}
      \put(4.8,3.0){\tiny $(b)$}
      \put(5.9,3.0){\tiny $(c)$}
      \put(7.0,3.0){\tiny $(d)$}
    \end{picture}
  \end{center}
  \caption{The chiral condensate as a function of $T/T_c$.}
  \label{ccc_T}
\end{figure}

\begin{figure}
  \unitlength1.0cm
  \begin{center}
    \begin{picture}(8.5,7.0)
      \put(1.0,0.5){
        \epsfysize=7.5cm
        \epsfxsize=6.0cm
        \rotate[r]{\epsffile{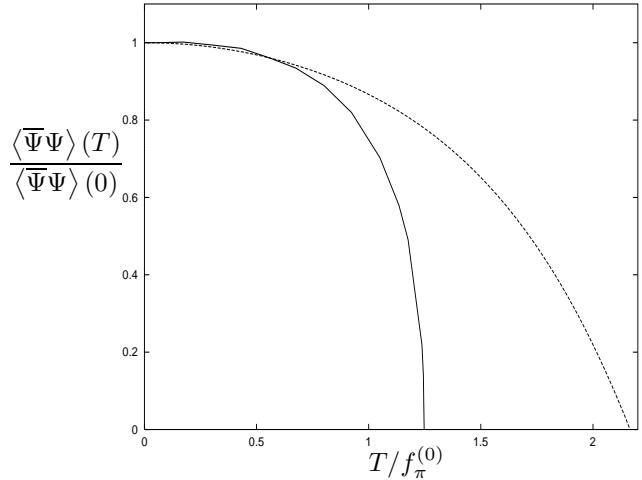}}
        }
      \put(0.0,4.2){\bf 
        $\ds{\frac{\VEV{\overline{\Psi}\Psi}(T)}
          {\VEV{\overline{\Psi}\Psi}(0)} }$}
      \put(4.8,0.2){\bf $\ds{T/f_\pi^{(0)}}$}
    \end{picture}
  \end{center}
  \caption{\footnotesize The chiral condensate
    $\VEV{\overline{\Psi}\Psi}$ as a function of $T/f_\pi^{(0)}$ in
    comparison with the result of chiral perturbation theory. Here
    $f_\pi^{(0)}$ denotes the pion decay constant for $\hat{m}=0$.}
  \label{cc_T}
\end{figure}

The introduction of finite temperature in the exact RG formalism is
straightforward. Since we are working in $4d$ Euclidean space time, at
finite $T$ the $4d$ momentum integral implicitly contained in the trace
in~(\ref{ERGE}) is replaced in the standard way by a $3d$ integration times
a Matsubara sum. Consequently, all scalar and fermion degrees of freedom
are substituted by infinite towers of Matsubara modes with increasing
$T$--dependent masses. Technically, one can treat $T$ as an external
parameter and evolve the effective average action (\ref{GammaEffective})
for each value of $T$ from $k_\Phi$ to $k=0$ using~(\ref{ERGE}).  In this
context it is important that the initial conditions at $k_\Phi$ for the RG
flow discussed above are practically insensitive to $T$--effects up to
temperatures of approximately $170\MeV$~\cite{BJW98-1}.  It is crucial that
the decoupling of massive modes through threshold functions also works for
Matsubara masses.  For instance, eq.~(\ref{AAA85}) is now replaced by
\begin{eqnarray}
 \label{AAA85b}
 \ds{l_n^4(\frac{M^2}{k^2},\frac{T}{k};\eta_\Phi)} &=& \ds{8n\pi^2
   k^{2n-4}T\sum_{l\in\ZZZ}
   }\nnn
 &\times& \ds{
   \int\frac{d^3\vec{q}}{(2\pi)^4}
   \frac{\partial_t(Z_\Phi^{-1}R_k(q^2))}
   {\left[ P(q^2)+M^2\right]^{n+1}}
   }
\end{eqnarray}
where
\begin{equation}
  \label{AAA85c}
  q^2=(2l\pi T)^2+\vec{q}^2\; .
\end{equation}
Using $P(q^2)\equiv q^2+Z_\Phi^{-1}R_k(q^2)$ we see that the squared scalar
mass $M^2$ is effectively replaced in~(\ref{AAA85b}) by the Matsubara
masses $M^2+(2l\pi T)^2$, $l\in\ZZ$.  The threshold functions therefore
automatically decouple all massive Matsubara modes as $k$ is lowered and
the model is dimensionally reduced in a smooth way.  This procedure yields,
for instance, the full effective potential as a function of $\Phi$, $T$ and
the external source $\jmath$ and therefore the equation of state $\partial
U(\VEV{\Phi},T)/\partial\Phi=\jmath$ which contains the $T$--dependence of
various observables.  In fig.~\ref{ccc_T} we show the chiral condensate
$\VEV{\overline{\Psi}\Psi}$ as a function of $T/T_c$.  Lines $(a)$, $(b)$,
$(c)$, $(d)$ correspond to $m_\pi(T=0)=0,45\MeV,135\MeV,230\MeV$,
respectively. For each pair of curves the lower one represents the full
$T$--dependence of $\VEV{\overline{\Psi}\Psi}$ whereas the upper one shows
for comparison the universal scaling form of the equation of state for the
$O(4)$ Heisenberg model. We see a sharp second order phase transition in
the chiral limit (this claim will be substantiated further below by
demonstrating the appropriate critical behavior near $T_c$).
For non--vanishing quark masses this changes into a smooth crossover which
becomes smoother for increasing $\hat{m}$. The values for the respective
(pseudo)critical temperatures are given in table~\ref{tab11}.  It is
remarkable that $T_{c}$ for the largest pion mass, computed within this
very simple effective model, is quite close to the result of current full
two flavor QCD Monte Carlo simulations performed for comparable quark
masses~\cite{LP97-1}. We consider this as an indication that our model
yields indeed even quantitatively a reasonable description of the chiral
transition. This hold for non--universal as well as for universal
properties of the transition as will be demonstrated below.  This claim is
further substantiated by the following consideration: Because of the
relatively large sigma mass~\cite{JW96-1,BJW98-1}, for low temperatures the
linear quark meson model is well approximated by the nonlinear sigma model.
In fact, our $\VEV{\overline{\Psi}\Psi}(T)$ curve rather accurately agrees
with that of chiral perturbation theory for $T\lta60\MeV$ as shown in
figure~\ref{cc_T}.  Here the solid line displays the exact RG result for
vanishing average current quark mass $\hat{m}=0$ whereas the dashed line
shows the corresponding result of three--loop chiral perturbation
theory~\cite{GL87-1,Leu88-1} On the other hand, in the vicinity of $T_c$
the universal behavior of the model sets in which is independent of the
details of the effective action used in our approach and also accurately
described by our method as demonstrated below.
\begin{table}
  \begin{center}
    \caption{\footnotesize The table shows the critical and
      ``pseudocritical'' temperatures for various values of the zero
      temperature pion mass.}
    \begin{tabular}{c||c|c|c|c}
      $\stackrel{ }{\frac{m_\pi}{\stackrel{\MeV}{  }}}$ &
      $0$ &
      $45$ &
      $135$ &
      $230$
      \\ \hline\hline
      $\stackrel{ }{\frac{T_{p c}}{\stackrel{\MeV}{  }}}$ &
      $100.7$ &
      $\simeq110$ &
      $\simeq130$ &
      $\simeq150$
      \\
    \end{tabular}
    \label{tab11}
  \end{center}
\end{table}

\begin{figure}
  \unitlength1.0cm
  \begin{center}
    \begin{picture}(8.5,7.0)
      \put(1.0,0.5){
        \epsfysize=7.5cm
        \epsfxsize=6.0cm
        \rotate[r]{\epsffile{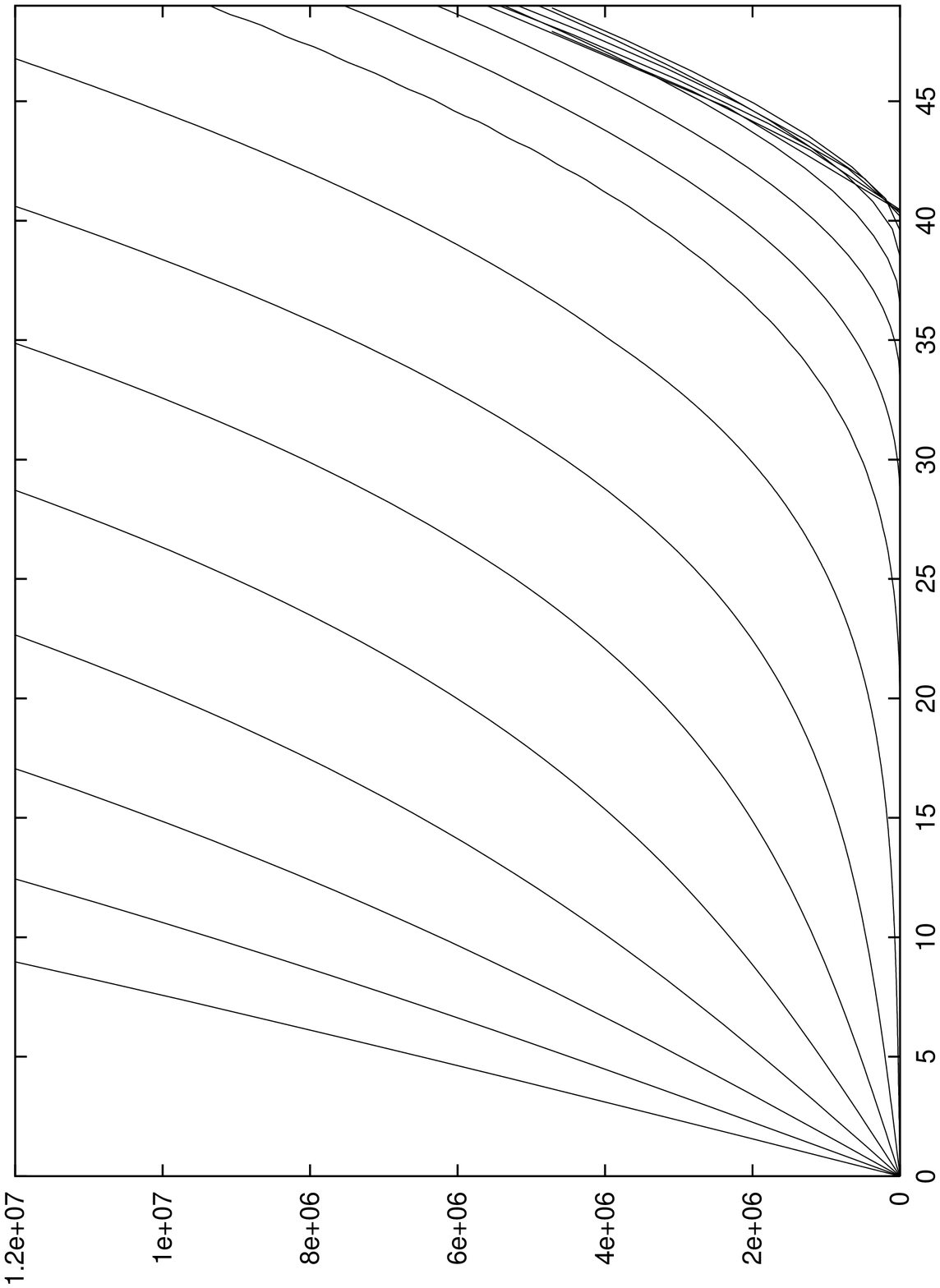}}
        }
      \put(0.2,4.0){\bf $\ds{\frac{\frac{U(T)}{\partial \phi_R}}
          {\MeV^{3}}}$}
      \put(4.3,0.2){\bf $\phi_R/\MeV$}
    \end{picture}
  \end{center}
  \caption{\footnotesize The plot shows the derivative of the
    meson potential $U(T)$ with respect to the renormalized field
    $\phi_R=(Z_\Phi\rho/2)^{1/2}$ for different values of $T$.
    }
  \label{Usig}
\end{figure}

\begin{figure}
  \unitlength1.0cm
  \begin{center}
    \begin{picture}(8.5,7.0)
      \put(1.0,0.5){
        \epsfysize=7.5cm
        \epsfxsize=6.0cm
        \rotate[r]{\epsffile{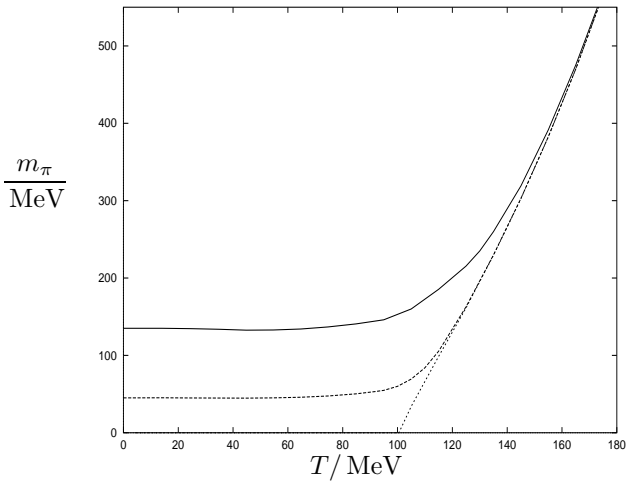}}
        }
      \put(0.2,4.0){\bf $\ds{\frac{m_\pi}{\MeV}}$}
      \put(4.3,0.2){\bf $\ds{T/\MeV}$}
    \end{picture}
  \end{center}
  \caption{\footnotesize $m_\pi$ as a function of $T$.}
  \label{mpi_T}
\end{figure}

\begin{figure}[t]
  \unitlength1.0cm
  \begin{center}
    \begin{picture}(8.5,7.0)
      \put(1.0,0.5){
        \epsfysize=7.5cm
        \epsfxsize=6.0cm
        \rotate[r]{\epsffile{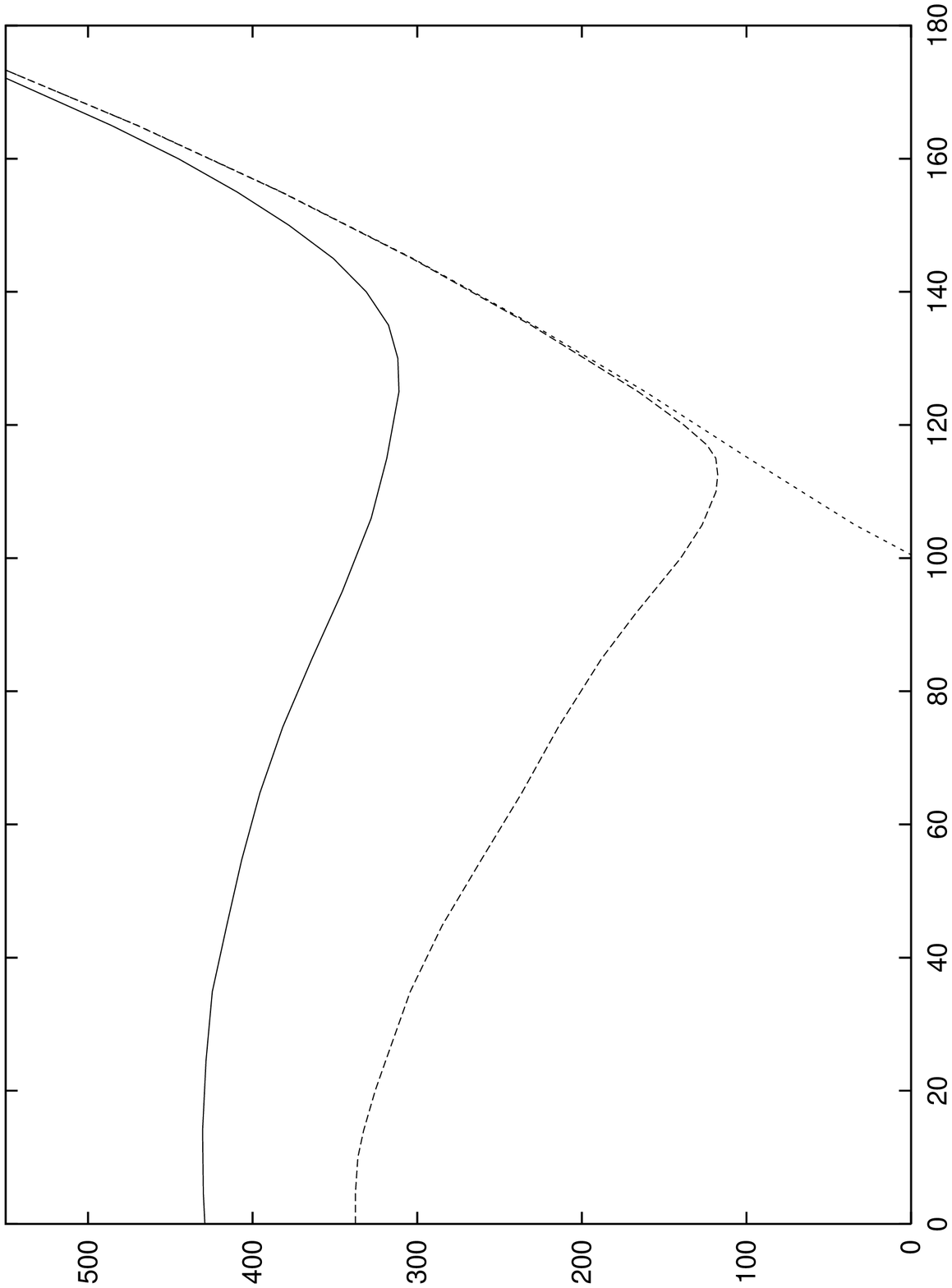}}
        }
      \put(0.2,4.0){\bf $\ds{\frac{m_\si}{\MeV}}$}
      \put(4.6,0.2){\bf $\ds{T/\MeV}$}
      \put(5.9,1.2){\footnotesize\bf $m_\pi=0$}
      \put(2.8,4.1){\footnotesize\bf $m_\pi=45\MeV$}
      \put(2.8,5.3){\footnotesize\bf $m_\pi=135\MeV$}
    \end{picture}
  \end{center}
  \caption{\footnotesize $m_\si$ as a function of $T$.}
  \label{ms_T}
\end{figure}

In order to demonstrate our ability to compute the complete temperature
dependent effective meson potential we plot in figure~\ref{Usig} $\partial
U(T)/\partial\phi_R$ as a function of the renormalized field variable
$\phi_R=(Z_\Phi\rho/2)^{1/2}$ for different values of $T$.
The first curve on the left corresponds to $T=175 \MeV$. The successive
curves to the right differ in temperature by $\Dt T=10 \MeV$ down to $T=5
\MeV$. One nicely observes the convexity of the potential even deep in
the spontaneously broken phase, i.e. for small enough temperatures.  This
property of $U$ is rather difficult to reproduce correctly in perturbation
theory.
The effective potential $U(\phi_R,T)$ also contains the necessary
information to compute the renormalized masses (or rather inverse spatial
correlation lengths) of the pions and the sigma meson. These are obtained
from $U$ as second derivatives with respect to the appropriate scalar field
components. Our results for the $T$--dependence of these two quantities are
plotted in figures~\ref{mpi_T} and~\ref{ms_T}, respectively, for three
different values of $M_\pi(T=0)$.
The solid lines in both cases correspond to the realistic value
$\hat{m}=\hat{m}_{\rm phys}$ whereas the dotted line represents the
situation without explicit chiral symmetry breaking, $\hat{m}=0$.  The
intermediate dashed lines are computed for $\hat{m}=\hat{m}_{\rm phys}/10$.
We note that, even for the case of the physical current quark mass, $m_\pi$
is to a good approximation a monotonically increasing function of $T$ for
all three quark masses. This implies, in particular, that for two light
quark flavors there is no long pion correlation length in thermal
equilibrium.  This is interesting in view of recent
speculations~\cite{RaWi93-1,Raj95-1} that such long correlation lengths
might lead to spectacular experimental signatures in heavy ion collision
experiments due to the formation of large regimes of a disoriented chiral
condensate.  A further interesting result is the dip in the $T$--dependence
of the sigma mass. For vanishing temperature the sigma decays predominantly
into two pions making it extremely wide through the large scalar
self--interaction. From figure~\ref{ms_T} we see that for increasing
temperature the phase space for this decay decreases until at temperatures
around $\sim100\MeV$ this decay becomes kinematically impossible. One would
therefore expect to see a strong $T$--dependence of the sigma decay widths
at least in thermal equilibrium.

\section{Universal critical behavior}
\label{sec5}

In addition to the non--universal results discussed above also universal
properties of the second order phase transition are described accurately by
the exact RG method.  Using the equation of state allows to substantiate
the claim that the chiral transition in the chiral limit $\hat{m}=0$ is
indeed of second order and to compute the associated critical exponents.
Table~\ref{tab2} shows our results which correspond to the critical
exponents of the three--dimensional $O(4)$--Heisenberg model.  Our results
are denoted by ``ERG'' whereas ``MC'' labels the exponents obtained by
lattice simulations \cite{KK95-1}. The agreement is within a few percent
except for the anomalous dimension. The latter deviation can be understood
at least qualitatively as a consequence of the rather crude approximation
of the momentum dependence of the scalar propagator
in~(\ref{GammaEffective}). In fact, if the first non--trivial order of the
derivative expansion in the scalar sector of the model is completed by
including in the Ansatz~(\ref{GammaEffective}) a term $\sim
Y_\Phi(\rho)\partial_\mu\rho\partial^\mu\rho$ the values of the critical
exponents for general $O(N)$--models can be seen to improve~\cite{TW94-1}.

\begin{figure}
  \unitlength1.0cm
  \begin{center}
    \begin{picture}(8.5,7.0)
      \put(0.0,-2.5){
        \epsfxsize=9.5cm
        \epsfysize=11.0cm
        {\epsfbox{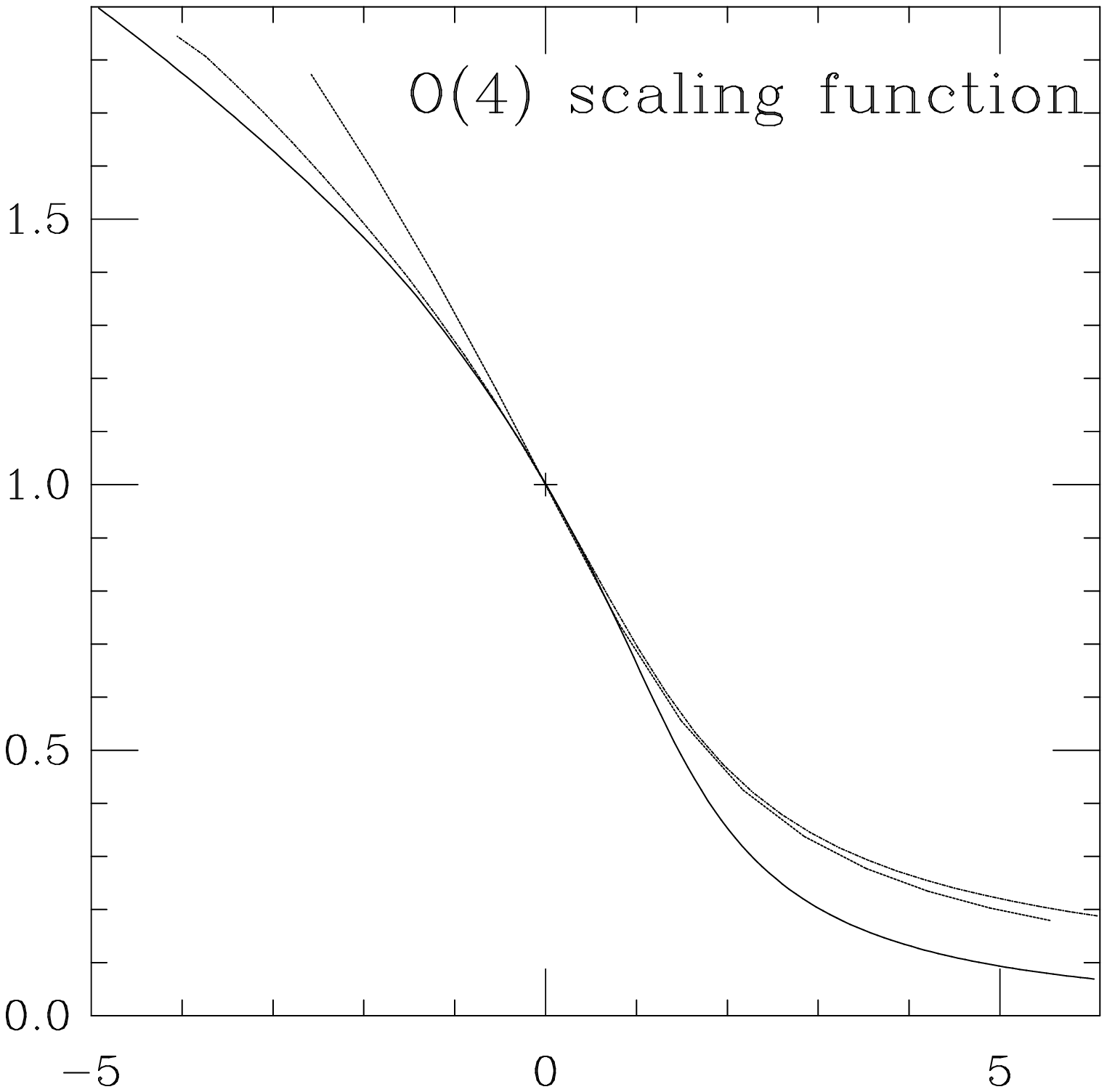}}
        }
      \put(1.20,0.83){
        \epsfysize=7.1cm
        \epsfxsize=5.9cm
        \rotate[r]{\epsffile{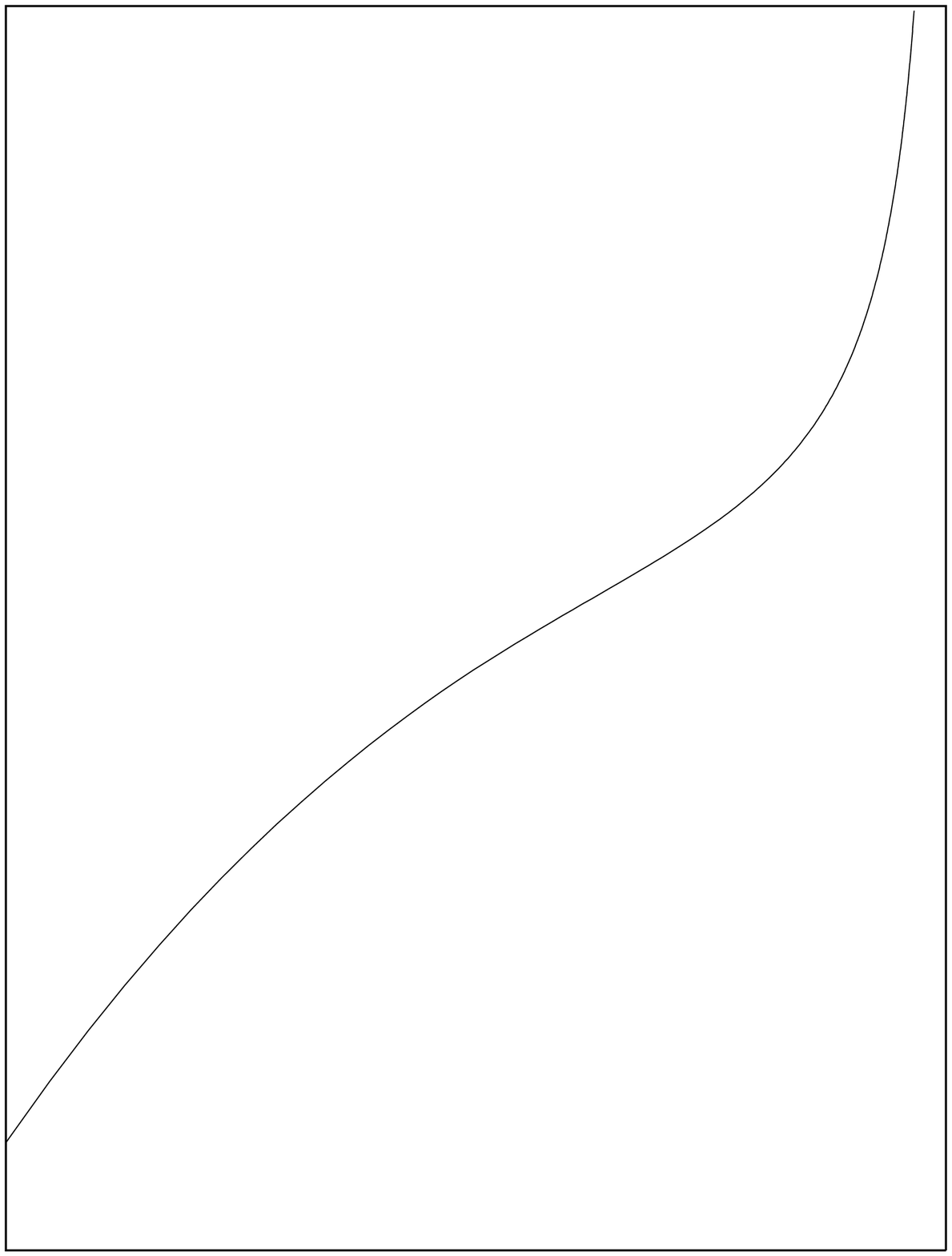}}
      }
      \put(4.3,0.2){\bf $\frac{(T-T_c)/T_c}
          {(\jmath/T_c^3 B^{\delta} D)^{1/\beta \delta}}$}
      \put(0.0,3.7){\bf $\frac{2\overline{\si}_0/T_c}
          {(\jmath/T_c^3 D)^{1/\delta}}$}
      \put(2.8,6.5){\tiny $\mbox{ERG}$}
      \put(2.7,5.7){\tiny $\mbox{MC}$}
      \put(2.4,6.5){\footnotesize $\epsilon$}
      \put(3.8,5.7){\tiny $\mbox{MF}$}
      \put(5.5,1.8){\tiny $\mbox{ERG}$}
      \put(6.5,1.7){\tiny $\mbox{MC}$}
      \put(6.9,2.0){\footnotesize $\epsilon$}
      \put(7.8,1.5){\tiny $\mbox{MF}$}
    \end{picture}
  \end{center}
  \caption{The critical equation of state for the $3d$ $O(4)$--Hei\-sen\-berg
    model.}
  \label{fig2}
\end{figure}

Yet, the exact RG method is even capable of computing the full Widom
scaling form of the equation of state.  In fig.~\ref{fig2} a comparison of
our results, denoted by ``ERG'', with results of other methods for the
scaling function of the three--dimensional $O(4)$ Heisenberg model is shown
(for notation see~\cite{BJW98-1}). The constants $B$ and $D$ specify
non--universal amplitudes of the model.  The curve labeled by ``MC''
represents a fit to lattice Monte Carlo data~\cite{Tou97-1}.
The second order epsilon expansion~\cite{BWW73-1} and
mean field results are denoted by ``$\epsilon$'' and ``MF'', respectively.
Apart from our results the curves are taken from~\cite{Tou97-1}.
\begin{table}
  \begin{center}
    \caption{Critical exponents of the $3d$ $O(4)$--model}
    \label{tab2}
    \begin{tabular}{c l l l l l }
      &
      $\nu$ &
      $\gamma$ &
      $\delta$ &
      $\beta$ &
      $\eta$
      \\[0.5mm] \hline\hline
      ERG &
      $0.787$ &
      $1.548$ &
      $4.80$ &
      $0.407$ &
      $0.0344$
      \\ \hline
      MC &
      $0.7479(90)$ &
      $1.477(18)$ &
      $4.851(22)$ &
      $0.3836(46)$ &
      $0.0254(38)$
      \\
    \end{tabular}
  \end{center}
\end{table}\nopagebreak

We should emphasize, perhaps, that these universal results by no means
prove that two flavor QCD is in the universality class of the $3d$
$O(4)$--symmetric Heisenberg model. As we pointed out before there is the
possibility that the axial $U_A(1)$ symmetry might be effectively restored
at high temperatures in which case one would rather expect a $U_L(2)\times
U_R(2)$--symmetric behavior near the chiral transition. Our results rather
confirm that the exact RG method is capable of computing highly
nonperturbative quantities like critical exponents or the scaling form of
the equation of state with good accuracy. Assuming that the axial $U_A(1)$
symmetry remains sufficiently broken even at high $T$ we would then expect
our results to be in good qualitative and quantitative agreement with the
universal as well as the non--universal properties of the chiral phase
transition of full two flavor QCD. This is based on the fact that our
method provides a smooth link between two temperature regimes where the
properties of strongly interacting matter are well understood:
\begin{itemize}
\item low temperatures where chiral perturbation theory is expected to
  yield a sound description of the dynamics of the Goldstone modes
\item temperature in the vicinity of $T_c$ where the universal critical
  behavior of the model sets in and the system becomes independent of almost
  all the details of the underlying microscopic description.
\end{itemize}
The plots~\ref{ccc_T} and~\ref{cc_T} demonstrate that the intermediate
range of temperature, where in principle sizeable quantitative deviations
from full two flavor QCD are possible, is not very large.

\acknowledgments

It is a pleasure to thank the organizers of the {\em 5th International
  Workshop on Thermal Field Theories and Their Applications} for having
provided a most stimulating conference environment. I would also like to
express my gratitude to J.~Berges and C.~Wetterich for their collaboration
on the topics presented here.


\end{document}